# Can we predict the mutation rate at the single nucleotide scale in the human genome?

Adam Eyre-Walker*
Ying Chen Eyre-Walker

School of Life Sciences
University of Sussex
Brighton
BN1 9QG

*Correspondence: a.c.eyre-walker@sussex.ac.uk

**Abstract**

It has been recently claimed that it is possible to predict the rate of de novo mutation of each site in the human genome with almost perfect accuracy (Michaelson et al. (2012) Cell, 151, 1431-1442). We show that this claim is unwarranted. By considering the correlation between the rate of *de novo* mutation and the predictions from the model of Michaelson et al., we show that there could be substantial unexplained variance in the mutation rate. We also demonstrate that the model of Michaelson et al. fails to capture a major component of the variation in the mutation rate, that which is local but not associated with simple context.

**Article**

It has been known for some time, from comparative studies, that the mutation rate varies at a number of different scales along the human genome, from variation between individual nucleotides, to differences between whole chromosomes (Hodgkinson and Eyre-Walker, 2011). Much of this variation has remained unexplained (Hodgkinson and Eyre-Walker, 2011). However, Michaelson et al. (Michaelson, et al., 2012) have recently claimed that the rate of mutation at each site is almost perfectly predictable. They use principle component logistic regression fitted to a dataset of 653 *de novo* mutations (DNMs) to estimate a model from which they can predict the mutation index (MI), a measure of the mutation rate, of each site in the human genome. To assess the fit of the model they count the number of sites in the genome with a particular MI (*n*) and the number of DNMs at those sites (*d*). They therefore have a prediction of the mutation rate from their model, the MI, and the observed rate of mutation, *z=d/n*. They find a very strong correlation between the logarithm of *z* and MI and infer that their model explains >90% of the variance in mutation rates. However, for each MI value they have thousands to millions of sites. As a consequence any variation that their model does not explain will tend to be averaged out when they consider the observed number of mutations. This can be illustrated as follows. Consider sites with an MI such that their mutation rate is $10^{-8}$, approximately the mean mutation rate in humans (1000_Genomes_Project_Consortium, 2010; Awadalla, et al., 2010; Conrad, et al., 2011). If the model of Michaelson et al. (Michaelson, et al., 2012) explains all the variation in the mutation rate then all sites with this MI will have a mutation rate of $10^{-8}$. However, if there is unexplained variance the mutation rate of each site will deviate from this value. Let us assume that equal numbers of sites with this MI have mutation rates of $0.1 \times 10^{-8}$ and $1.9 \times 10^{-8}$. It is clear that if we only sample a few sites then the observed mutation rate will often deviate substantially from the expected value and the correlation between the log of the observed number of DNMs and the MI will be correspondingly weak. However, as we sample more and more sites so the mean value will approach the expected value of $10^{-8}$ and the correlation between the log of the number of DNMs per site and MI will become better (assuming that the model of Michaelson et al. explains at least some of the variance). Since there are typically thousands if not millions of sites for each MI value, any unexplained variance will be averaged out of sight.

We can estimate how much variance might be left unexplained by the model of Michaelson et al. (Michaelson, et al., 2012) (henceforth referred to as the Michaelson model) by simulating data under their model with and without additional variance. In the simulation we estimate the relationship between MI and the rate of mutation using a sets of DNMs. We then use this relationship to predict the expected number of mutations at a site and then simulate data based on these expectations (details in supplementary information).We performed the analysis for three sets of DNMs: (i) the 652 DNMs reported by Michaelson et al. (Michaelson, et al., 2012) and used to build the model upon which the MI values are based (referred to as the Michaelson data), (ii) 1380 DNMs reported by various other studies (Conrad, et al., 2011; Iossifov, et al., 2012; Neale, et al., 2012; O'Roak, et al., 2011; Sanders, et al., 2012)(Other data), and (iii) 4933 DNMs reported by Kong et al. (Kong, et al., 2012)(Kong data)(note that only DNMs with an MI value were included).

As previously shown by Michaelson et al. (Michaelson, et al., 2012), the correlation between the log of the number of DNMs per site and the MI value is very strong for the Michaelson data (r = 0.98, p<0.001; Figure 1a); this is perhaps not surprising given that this was the data used to construct the Michaelson model and the model is parameter rich. However, as Michaelson et al. (Michaelson, et al., 2012) showed, their model also fits the data from other studies well (r = 0.97, p<0.001; Figure 1b), although there is a clear non-linearity in the relationship (a quadratic term in a non-linear regression is significant p = 0.010). However, the fit of the Michaelson model to the Kong data, which Michaelson et al. did not study, is relatively poor (r = 0.94, p<0.001; Figure1c). The problem would seem to lie with the Kong data, since the model fits the other two datasets well. The slope of the regression line from the Kong data (0.0047 (0.0006)) is significantly less than that observed for the Michaelson et al. (0.010 (0.0007)) and other datasets (0.0084 (0.0007)) suggesting that there has been systematic under-reporting of DNMs from the more mutable areas of the genome in the Kong et al. dataset (or alternatively, that there are large numbers of false positives in the less mutable parts).

If we assume that the Michaelson model explains all the variation in the mutation rate, we find that simulated datasets have similar levels of correlation, between the log of the number of DNMs per site and MI, to that observed in the real data for the Michaelson and other datasets; almost all the simulated correlations are stronger than the observed correlation in the Kong data, but this is probably because the Michaelson model clearly fits this data poorly. However, despite the good fit between model and data for two of the datasets, we find that there could be very substantial levels of unexplained variance and the correlations would remain almost unaffected. Only when the variance associated with the unexplained variance approaches $10^5$ do we see the correlations being affected and approaching the values seen in the real data. This level of variance dwarfs that explained by the Michaelson model; the coefficient of variation in the mutation rate explained by the Michaelson model is 1.10, the coefficient of variation for the unexplained variation is 300 if variance is $10^5$. This analysis therefore shows that there could be a substantial amount of unexplained variance that would never be detected assessing model fit as Michaelson et al. have done.

Assessing model fit is not easy within these datasets; there are very few DNMs spread across millions of sites. We therefore sought to test one component of mutation rate variation that is both substantial and likely to be difficult to predict, so called cryptic variation in the mutation rate (Hodgkinson, et al., 2009; Johnson and Hellmann, 2011). This is variation at the single nucleotide level that is independent of local sequence context. It has been estimated that there might be as much variation that is independent of context, as depends upon context (Hodgkinson, et al., 2009). The evidence for this so called "cryptic" variation comes from the observation that there is an excess of orthologous sites at which humans and chimpanzees have a SNP (Hodgkinson, et al., 2009; Johnson and Hellmann, 2011), and an excess orthologous sites at which there is a substitution between human and chimpanzee, and a substitution between orangutan and rhesus macaque (Johnson and Hellmann, 2011). The excess of coincident SNPs cannot be explained by ancestral polymorphism, natural selection or sequencing problems (Hodgkinson, et al., 2009; Johnson and Hellmann, 2011). It therefore appears that the excess of coincident SNPs, and substitutions at identical positions in different species, is due to variation in the mutation rate.

To investigate whether the model of Michaelson et al. captures cryptic variation in the mutation rate we proceeded as follows. Leffler et al. (Leffler, et al., 2013) have shown, using a carefully curated dataset of human and chimpanzee SNPs, that there is a 16% excess of coincident SNPs at CpG sites (95% confidence intervals of 14% and 17%) and a 83% (80%, 86%) excess at non-CpG sites between human and chimpanzee. We can use some theory set out Hodgkinson et al. (Hodgkinson, et al., 2009) to infer how much variation in the mutation rate is consistent with this excess of coincident SNPs and then to estimate the average mutation rate of coincident SNPs relative to the genomic average (see supplementary material for details). We estimate that sites with coincident SNPs are 1.4x (1.4x, 1.4x) and 2.7x (2.7x, 2.8x) more mutable than the genomic average for CpG and non-CpG sites respectively. How do these values compare to those under the Michaelson model? Under the Michaelson model we find that sites with coincident SNPs have significantly greater MI values at both CpG (mean MI for coincident sites = 91.6, non-coincident sites = 81.4; $p < 0.001$) and non-CpG sites (coincident sites = -7.77, non-coincident sites = -16.0, $p<0.001$). However, the differences in MI are small and equate to minor differences in the mutation rate predicted using the regression model from the Michaelson et al. data; coincident SNPs are predicted to be 27% more mutable at CpG and 21% more mutable at non-CpG sites. Thus our analysis suggests that the Michaelson model captures much of the variation at CpG sites; the level of variation required to explain the excess of coincident SNPs at CpG sites is such that we would expect sites with coincident SNPs to be 40% more mutable than non-coincident sites and the Michaelson model predicts them to 27% more mutable. However, the Michaelson model seems to fail to capture much of the variation at non-CpG sites; sites with coincident SNPs are expected to be 270% more mutable than average sites, but the Michaelson model predicts them to be only 21% more mutable.

The Michaelson model clearly captures some of the variation in the mutation rate, but how much of the variation is far from clear. It does not appear to capture variation in the mutation rate at non-CpG sites, which is independent of context, but the contribution of this variation to the overall variance in the mutation is also still unknown.

**Acknowledgements**

The authors are grateful to Jake Michaelson and Peter Keightley for helpful discussion.

## Supplementary material

**Materials and Methods**

*Data*

*Simulating data*

We simulated data as follows under the model of Michaelson et al. (Michaelson, et al., 2012) as follows. First, for a dataset of DNMs we regressed, using weighted regression, the log of the observed number of DNMs per site, $z$, against MI, to yield the relationship between the mutation rate and MI under the Michaelson model. Since there are a limited number of DNMs for some MI values we binned the MI values into groups of ten, and removed those bins that had 5 or fewer DNMs. Using the regression equation, and the number of sites, we predicted the expected number of mutations at sites with an MI of $x$, $Z(x)$. To generate data under the assumption that the Michaelson model explains all the variance in the mutation rate we sampled from a Poisson distribution with expected values $Z(x)$. To investigate the effect of variance unexplained by the Michaelson model we added an additional step to the simulation. Having used the regression model (of log(DNMs per site) versus MI) to predict the expected number of mutations for a site with an MI of $x$, $Z(x)$ we multiplied this by a random variate drawn from a lognormal distribution with variance = $v/n$, where $n$ is the number of sites, befoe sampling from a Poisson distribution. The logic is as follows; the mean mutation rate for sites with an MI of $x$ is $Z(x)$, but the rate of a particular site is $Z(x)\alpha$ where $\alpha$ is a random variate that is lognormally distributed. Since, the mean of $n$ lognormally distributed variates, each with a variance $v$, is itself approximately lognormal with a variance equal to $v/n$ (Beaulieu, et al., 1995; Fenton, 1960), we can simulate the effect of unexplained variation amongst sites with an MI of $x$ by multiplying the expected mutation rate by a random lognormal variate with variance $v/n$. We generated 1000 simulated datasets and calculated the correlation between MI and the log of the simulated number of mutations per site. Occasionally the simulation would generate no DNMs for an MI value; we removed these datasets. We then compared the correlation between the log of the observed number of DNMs and MI, against the correlation between the log of the simulated number of DNMs and MI. To take into account the uncertainty in the relationship between the log of the observed mutation rate and MI, we bootstrapped the data prior to performing the regression by resampling the datapoints from the regression.

*Coincident SNP calculation*
We investigated the difference in the mutation rate between sites with and without a coincident SNP as follows. We assume that the distribution of mutation rates is a gamma distribution arbitrarily scaled such that the mean of the distribution is one; it is therefore characterized solely by its shape parameter. We also assume that hypermutable sites destroy themselves when they mutate; this seems the most likely model. This assumption makes little difference to the non-CpG analysis, but reduces the level of variation needed to explain the coincident SNPs in the CpG analysis. Hodgkinson et al. (Hodgkinson, et al., 2009) have shown that under this model the probability of observing a coincident SNP at a site is

$$P = u_h u_c \int D(\gamma)(e^{-v\gamma}\gamma^2 + (1-e^{-2v\gamma}))d\gamma \qquad (1)$$

where $u_h$ and $u_c$ are the density of SNPs in the two species being considered, $v$ is the average divergence between the species and $D(\gamma)$ is the distribution of the rates. Therefore the average mutation rate of sites with coincident SNPs, relative to the average mutation rate (arbitrarily set to one) is

$$Q = \frac{\int \left(D(\gamma)e^{-v\gamma}\gamma^2 + (1-e^{-2v\gamma})\right)\gamma d\gamma}{\int \left(D(\gamma)e^{-v\gamma}\gamma^2 + (1-e^{-2v\gamma})\right)d\gamma} \qquad (2)$$

We assume that mutation rate is drawn from a gamma distribution. In our calculations we assume that the divergence at non-CpG sites between human and chimpanzee sites is 0.0092 (Chimpanzee-Sequencing-and-Analysis-Consortium, 2005) with the divergence at CpG sites 10x higher at 0.092 (Chimpanzee-Sequencing-and-Analysis-Consortium, 2005; Hwang and Green, 2004).

| *v* | Michaelson et al. | Other | Kong et al. |
|---|---|---|---|
| 0 | 0.81 | 0.78 | 1.0 |
| 1000 | 0.82 | 0.79 | 1.0 |
| 10,000 | 0.81 | 0.76 | 1.0 |
| 100,000 | 0.76 | 0.72 | 0.99 |
| 500,000 | 0.59 | 0.49 | 0.92 |
| 1,000,000 | 0.43 | 0.35 | 0.70 |
| 2,000,000 | 0.25 | 0.17 | 0.36 |
| 3,000,000 | 0.16 | 0.088 | 0.20 |
| 4,000,000 | 0.10 | 0.069 | 0.13 |
| 5,000,000 | 0.076 | 0.032 | 0.084 |

**Table 1.** The proportion of simulated datasets with a greater correlation between the log of the number of DNMs per site and MI, than observed in the actual data.

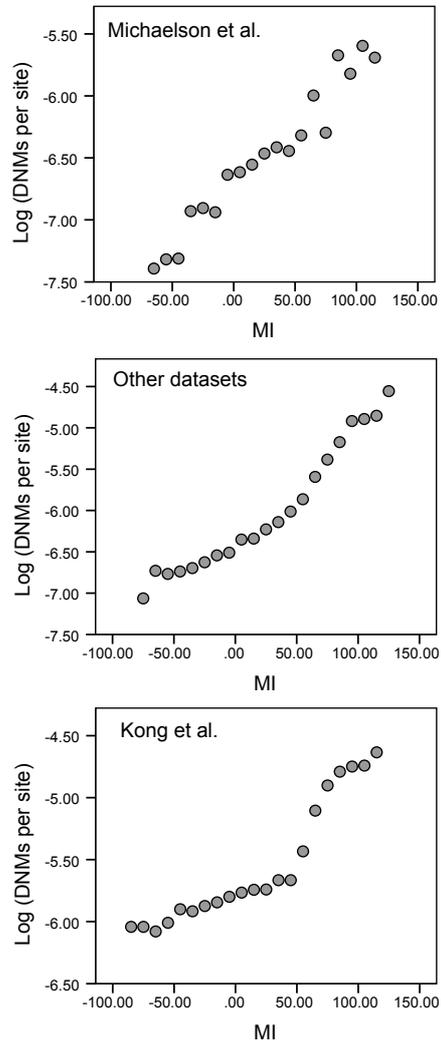

**Figure 1.** The log of the number of DNMs per site versus the mutation index.